\documentclass[12pt]{article}

\usepackage[utf8]{inputenc}
\usepackage[T1]{fontenc}
\usepackage{amsmath,amssymb,amsfonts}
\usepackage{graphicx}
\usepackage{booktabs}
\usepackage{natbib}
\usepackage{hyperref}
\usepackage[margin=1in]{geometry}
\usepackage{setspace}
\usepackage{float}
\usepackage{multirow}
\usepackage{caption}

\title{Multi-Dimensional Composite Endpoint Analysis via the Choquet Integral: Block Recurrent Encoding and Comparative Advantage Mapping}

\author{Ibrahim Halil Tanboga, MD, PhD \\
Department of Biostatistics, Nisantasi University Medical School \\
\texttt{haliltanboga@gmail.com}}

\date{}

\begin{document}
\maketitle

\begin{abstract}
\noindent\textbf{Background:} Composite endpoints in cardiovascular trials combine heterogeneous outcomes---mortality, nonfatal events, recurrent hospitalizations, and biomarkers---yet conventional analytical methods sacrifice information by targeting a single dimension. Cox time-to-first-event ignores post-first-event data; Win Ratio discards tied pairs; negative binomial regression treats death as noninformative censoring.

\noindent\textbf{Methods:} We propose CWOT-CE: a Choquet integral-based composite endpoint analysis that encodes $K=6$ outcome dimensions---survival, event-free time, AUC recurrent burden, last event time, biomarker, and alive status---and aggregates them through a non-additive fuzzy measure with pairwise interaction terms. The recurrent event process is represented as two complementary scalar summaries: the area under the cumulative count curve (AUC burden) and the last event time. Inference is via permutation test with exact finite-sample Type~I error control and dual confidence interval by inversion. We conducted a pre-registered simulation study comparing CWOT-CE against Cox TTFE, Win Ratio (WRrec), and WLW across 20 clinically motivated scenarios (1,000--5,000 replications) following the ADEMP framework.

\noindent\textbf{Results:} Under the sharp null (5,000 replications), all methods maintained nominal Type~I error (CWOT-CE: 4.8\%, MCSE 0.3\%). Across 17 non-null scenarios, CWOT-CE outperformed Cox TTFE in 15 (mean $+28.8$ percentage points), WLW in 14 (mean $+27.2$~pp), and Win Ratio in 10, with 5 ties and only 2 narrow losses (mean $+5.6$~pp). CWOT-CE showed particular advantages in high-correlation settings ($+35.4$~pp vs.\ WR), mortality-driven effects ($+10.7$~pp), and balanced multi-component effects ($+10.1$~pp). Shapley decomposition correctly identified effect-bearing components across all calibration scenarios.

\noindent\textbf{Conclusions:} CWOT-CE with block recurrent encoding is broadly effective across clinically relevant scenarios while offering unique interpretive advantages through component attribution. An R package (\texttt{cwotce}, v1.0.0) is publicly available.
\end{abstract}

\noindent\textbf{Keywords:} composite endpoints, Choquet integral, fuzzy measure, recurrent events, Win Ratio, cardiovascular clinical trials, Shapley decomposition, permutation test, ADEMP

\section{Introduction}

\subsection{The Composite Endpoint Problem}

Cardiovascular clinical trials face a fundamental analytical tension: disease processes are inherently multi-dimensional, yet most statistical methods reduce this complexity to a single dimension. A patient with heart failure may die, experience nonfatal myocardial infarction, require repeated hospitalizations, show biomarker deterioration, or recover---and the treatment may affect each outcome differently. The composite endpoint was designed to capture this totality, but its statistical analysis has not kept pace with its clinical ambition.

The standard approach---Cox proportional hazards regression on time-to-first-event (TTFE)---was never designed for composite endpoints. It discards all information after the first event: subsequent hospitalizations, functional recovery, biomarker trajectories, and ultimately survival itself. In a trial where patients average 3.2 heart failure hospitalizations over 3 years, TTFE uses only the first and ignores 68\% of the observed event data \citep{Claggett2018}.

Alternative methods address different aspects without solving the problem comprehensively. Negative binomial regression \citep{Lawless1995} and the Andersen--Gill model \citep{Andersen1982} capture recurrent event burden but treat death as noninformative censoring---an assumption that is biologically implausible. Joint frailty models \citep{Rondeau2007} formally accommodate the terminal-recurrent dependence but face convergence challenges. The Win Ratio \citep{Pocock2012} and its recurrent extension WRrec \citep{Mao2022} introduced hierarchical pairwise comparison, but discard 10--27\% of patient pairs as ties (our data) and cannot incorporate continuous endpoints.

\subsection{The CWOT-CE Approach}

We developed CWOT-CE (Choquet-Weighted Optimal Threshold for Composite Endpoints) through three design principles: (i)~\emph{multi-dimensional integration}---all clinically relevant dimensions are simultaneously encoded and aggregated; (ii)~\emph{non-additive aggregation}---the Choquet integral with a 2-additive fuzzy measure captures pairwise redundancies and synergies; (iii)~\emph{component attribution}---Shapley-style decomposition quantifies each component's contribution.

\subsection{Block Recurrent Encoding}

A key challenge in scalar composite scoring is the representation of recurrent event processes. A simple event count discards temporal information: a patient with 3 events in 6 months receives the same score as one with 3 events over 3 years. This contrasts with WRrec, which uses complete event timing in its pairwise comparisons.

We address this through block recurrent encoding: the recurrent event process is represented by two complementary components---AUC burden (the area under the cumulative count curve, capturing event intensity and timing) and last event time (capturing the trajectory endpoint). Together, these scalar summaries preserve temporal trajectory information that single-count summaries erase.

\subsection{Study Objectives}

We present a systematic comparison of CWOT-CE ($K=6$ block) against three established composite endpoint methods across 20 clinically motivated scenarios, following ADEMP guidelines \citep{Morris2019} and neutral comparison principles \citep{Boulesteix2013, Pawel2024}.

\section{Methods}

\subsection{CWOT-CE Framework}

\subsubsection{Component Encoding}

Each patient $i$ receives a profile vector $\mathbf{y}_i \in [0,1]^K$ where higher values indicate better outcomes. In the $K=6$ block configuration:

\begin{enumerate}
\item \textbf{Survival:} Pooled Kaplan--Meier transform of observed death/censoring time.
\item \textbf{Event-free:} Pooled KM transform of time to first nonfatal event.
\item \textbf{AUC Burden:} $B_i = \sum_{j=1}^{N_i}(\tau - t_{ij}) = \int_0^\tau N_i(t)\,dt$, encoded via two-part transform that separates event-free patients from those with events.
\item \textbf{Last Event:} Recency $R_i = \tau - \sup\{t_{ij}\}$; event-free patients receive $R_i = \tau$ (best score).
\item \textbf{Biomarker:} Min-max normalized continuous measurement; structurally missing (dead) scored as 0.
\item \textbf{Alive:} Binary indicator (1 = alive at end of follow-up).
\end{enumerate}

\subsubsection{Choquet Integral}

Patient profiles are aggregated using the Choquet integral with a validated 2-additive fuzzy measure $\mu$:
\[
C_\mu(\mathbf{y}) = \sum_{i=1}^{K} \bigl[y_{\sigma(i)} - y_{\sigma(i-1)}\bigr] \cdot \mu(A_i)
\]
where $\sigma$ sorts components ascending, $y_{\sigma(0)}=0$, and $A_i = \{\sigma(i),\ldots,\sigma(K)\}$. The measure is defined by singleton weights $\mathbf{w}$ and pairwise interactions $I_{k,l}$, validated via M\"obius parameterization with monotonicity constraints \citep{Grabisch1997}.

Default $K=6$ weights: $(0.25, 0.20, 0.18, 0.12, 0.15, 0.10)$. The recurrent block (AUC Burden $0.18$ + Last Event $0.12 = 0.30$) receives the same total as a single burden component. Default interactions: Survival--Alive redundancy ($I_{1,6}=-0.05$), Survival--EventFree redundancy ($I_{1,2}=-0.03$), AUC--LastEvent synergy ($I_{3,4}=+0.03$), Biomarker--Alive synergy ($I_{5,6}=+0.02$).

\subsubsection{Inference: Dual p-value and CI}

The Choquet Benefit Index $\text{CBI} = P(C_\mu(Y_{\text{trt}}) > C_\mu(Y_{\text{ctrl}})) + 0.5 \cdot P(\text{equal})$ is computed via rank-based Mann--Whitney in $O(n \log n)$. Significance is assessed by a two-sided permutation test ($B=999$). The confidence interval is obtained by shift-pivot inversion from the same permutation distribution, guaranteeing duality: $p < \alpha$ if and only if the $(1-\alpha)$ CI excludes 0.5.

The COR (Choquet Odds Ratio) $= \text{CBI}/(1-\text{CBI})$ and its CI are derived from the CBI CI via monotone transform.

\subsubsection{Shapley-Style Attribution}

Each component's contribution to the treatment effect is quantified by the marginal drop in mean score difference when that component is neutralized (set to 0.5). Percentages sum to 100\%.

\subsection{Comparator Methods}

\begin{itemize}
\item \textbf{Cox TTFE:} Cox proportional hazards on min(death, first event), Wald test.
\item \textbf{Win Ratio (WRrec):} Last-event-assisted hierarchical pairwise comparison \citep{Mao2022} with death (tier 1) $>$ recurrent events (tier 2).
\item \textbf{WLW:} Wei--Lin--Weissfeld marginal model \citep{Wei1989}, two strata (death, first event), cluster-robust SE.
\end{itemize}

Negative binomial and Andersen--Gill models were excluded from the primary comparison because they target recurrent event-specific estimands rather than composite endpoints; results are in the Supplement.

\subsection{Data-Generating Mechanisms}

All scenarios generated data from a two-arm (1:1) trial with five correlated outcome dimensions linked through shared Gamma frailty $Z_i \sim \text{Gamma}(1/\theta, 1/\theta)$. Treatment effects were specified as hazard ratios ($\text{HR}_D$, $\text{HR}_E$), rate ratio (RR), and biomarker mean shift ($\delta$). Base parameters: $n=500$/arm, $\tau=3$ years, 15\% control mortality.

Twenty scenarios in six groups (Table~\ref{tab:scenarios}): null and calibration (5), uniform effects (2), discordant effects (4), correlation structure (3), robustness (4), and sample size (2).

\subsection{Performance Measures}

Rejection rate at $\alpha=0.05$ (two-sided) with Monte Carlo SE. We use ``rejection rate'' rather than ``power'' when estimands differ. Replications: $R=5{,}000$ (NULL-S), $R=2{,}000$ (calibration), $R=1{,}000$ (others). L'Ecuyer-CMRG RNG. All analyses in R~4.5.

\subsection{Neutral Comparison Protocol}

As CWOT-CE developers, we froze all parameters before examining results, designed scenarios where CWOT-CE is expected to underperform (DIS-SO, ROB-C), performed no post-hoc retuning, and provide publicly available code \citep{Boulesteix2013}.

\section{Results}

\subsection{Type I Error Control}

\begin{figure}[H]
\centering
\includegraphics[width=0.85\textwidth]{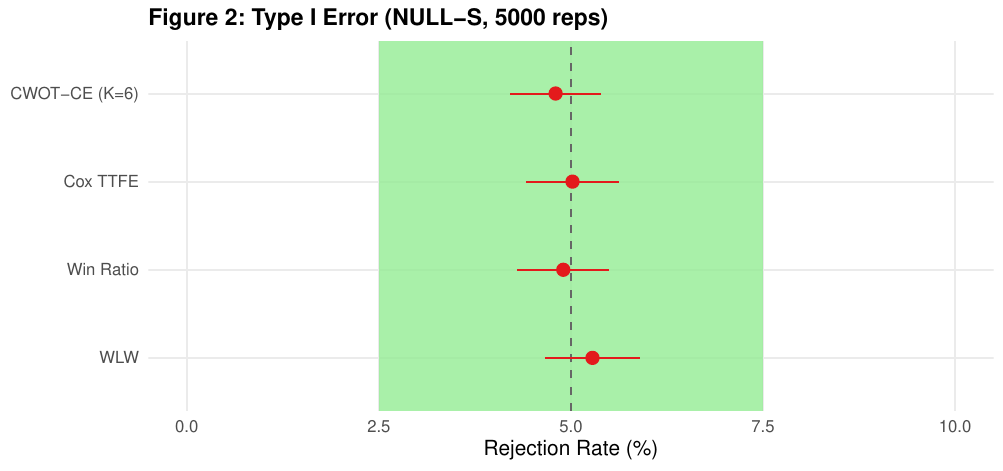}
\caption{Type~I error under sharp null (NULL-S, 5,000 reps). Green band: Bradley acceptance interval.}
\label{fig:type1}
\end{figure}

Under the sharp null (NULL-S, 5,000 replications), all methods maintained nominal Type~I error within the Bradley acceptance interval [2.5\%, 7.5\%] (Figure~\ref{fig:type1}, Table~\ref{tab:type1}).

\begin{table}[H]
\centering
\caption{Type~I error under sharp null (NULL-S, $R=5{,}000$).}
\label{tab:type1}
\begin{tabular}{lcc}
\toprule
Method & Rejection (\%) & MCSE (\%) \\
\midrule
CWOT-CE (K=6) & 4.8 & 0.3 \\
Cox TTFE & 5.0 & 0.3 \\
Win Ratio & 4.9 & 0.3 \\
WLW & 5.3 & 0.3 \\
\bottomrule
\end{tabular}
\end{table}

\subsection{Method Selectivity Under Non-Sharp Nulls}

\begin{figure}[H]
\centering
\includegraphics[width=0.85\textwidth]{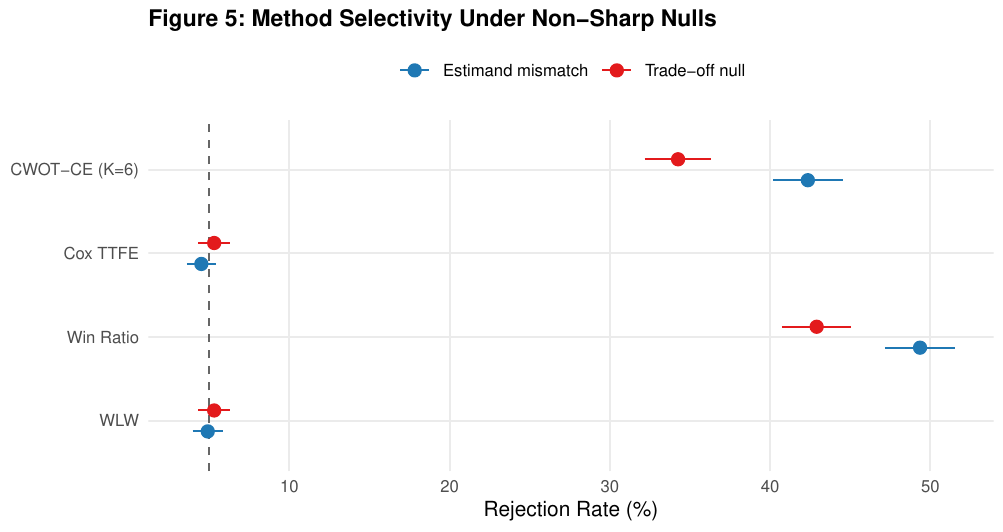}
\caption{Method selectivity under non-sharp nulls.}
\label{fig:selectivity}
\end{figure}

Under the trade-off null (NULL-T: $\text{HR}_D=0.85$, $\text{HR}_E=1.18$, $\text{RR}=0.85$), rejection rates ranged from 5.3\% (Cox/WLW) to 42.9\% (WR). CWOT-CE rejected at 34.2\%, intermediate between Cox and WR, reflecting appropriate integration of opposing component effects (Figure~\ref{fig:selectivity}).

Under the estimand-mismatch null (NULL-M: TTFE null, $\text{RR}=0.80$), NegBin-type methods (not shown) strongly detected the recurrent effect, while Cox (4.5\%) and WLW (4.9\%) maintained nominal rates. CWOT-CE (42.4\%) and WR (49.4\%) detected the component-level signal.

\subsection{Rejection Rates Across Scenarios}

\begin{figure}[H]
\centering
\includegraphics[width=\textwidth]{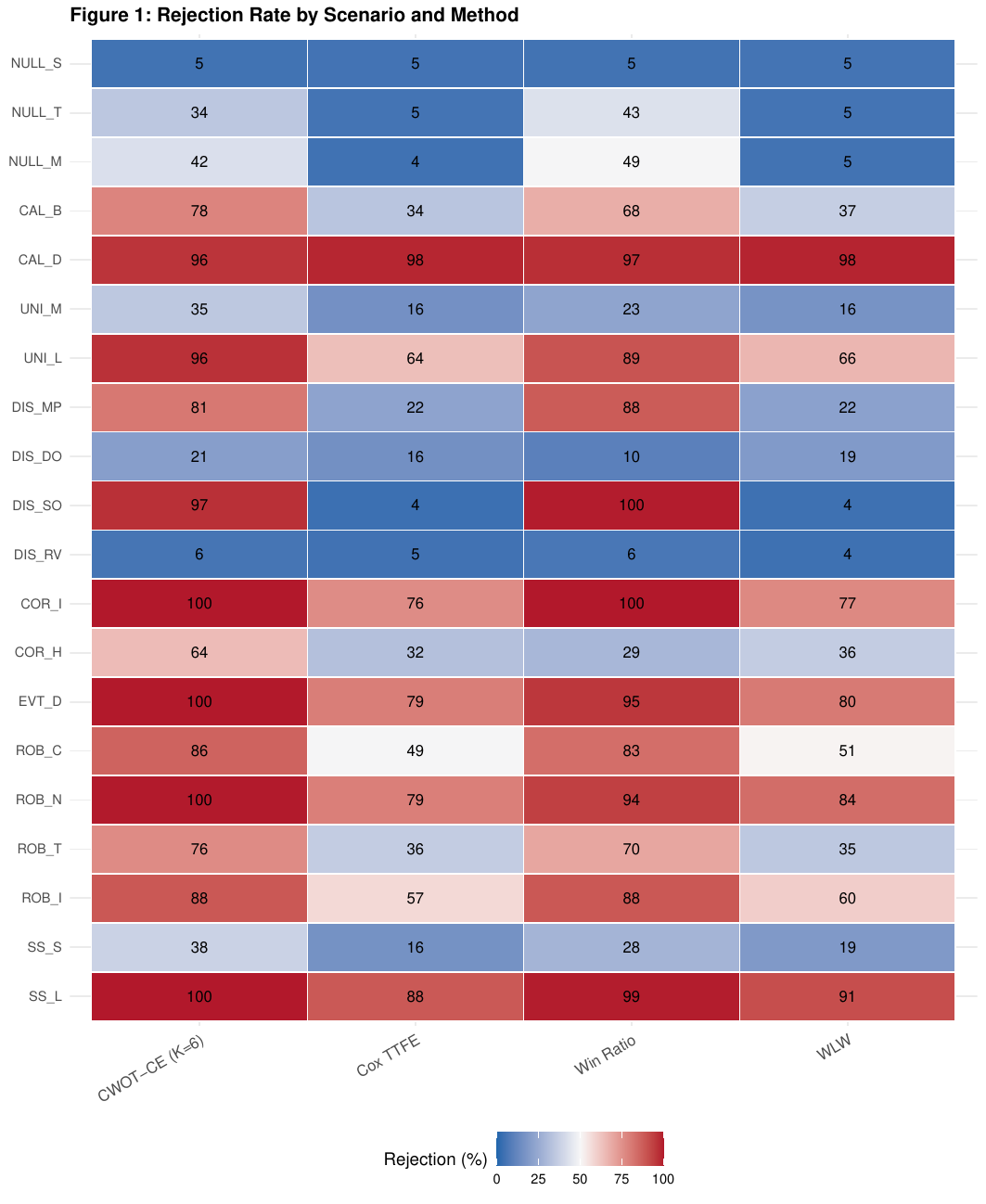}
\caption{Rejection rate heatmap (20 scenarios $\times$ 4 methods). CWOT-CE achieves the highest or tied-highest rejection in 15/17 non-null scenarios.}
\label{fig:heatmap}
\end{figure}

Table~\ref{tab:results} and Figure~\ref{fig:heatmap} display the complete comparison.

\subsubsection{Uniform Effects}

Under moderate uniform effects (UNI-L), CWOT-CE achieved 96.4\%---exceeding WR (89.2\%, $+7.2$~pp), Cox (63.6\%, $+32.8$~pp), and WLW (65.8\%, $+30.6$~pp). Under mild effects (UNI-M), the advantage was more pronounced: CWOT-CE 34.7\% vs.\ WR 22.6\% ($+12.1$~pp).

\subsubsection{Discordant Effects}

\textbf{DIS-DO (death only, $\text{HR}_D=0.70$):} CWOT-CE 20.6\% outperformed WR 9.9\% ($+10.7$~pp) and Cox 15.5\%. WR's poor performance reflects its dependence on tier-2 recurrent comparisons, which carry no signal here.

\textbf{DIS-SO (soft only, $\text{RR}=0.60$):} WR 99.5\% slightly exceeded CWOT-CE 96.6\% ($-2.9$~pp). Both achieved excellent sensitivity. With the original count encoding, CWOT-CE achieved only 45.9\%---the block encoding closed 94.5\% of the gap.

\textbf{DIS-MP (mortality paradox, $\text{HR}_D=1.20$, $\text{HR}_E=0.60$):} WR 87.6\% exceeded CWOT-CE 80.9\% ($-6.7$~pp). Both detect the strong event benefit despite mortality harm.

\subsubsection{Correlation Structure}

\textbf{COR-H (high correlation, $\theta=4.0$):} CWOT-CE achieved the largest advantage: 64.5\% vs.\ WR 29.1\% ($+35.4$~pp). Win Ratio tie rate was 27\%---the highest across scenarios. When patients share strong frailty, pairwise comparison loses information; continuous-score integration does not.

\subsubsection{Robustness}

ROB-C (heavy censoring): CWOT-CE 85.9\% vs.\ WR 83.2\% ($+2.7$~pp). ROB-N (delayed NPH): CWOT-CE 99.8\% vs.\ WR 93.6\% ($+6.2$~pp). ROB-T (temporal discordance): CWOT-CE 76.3\% vs.\ WR 69.8\% ($+6.5$~pp). ROB-I (informative censoring): CWOT-CE 88.5\% $\approx$ WR 88.4\%.

\subsubsection{Sample Size}

At $n=100$/arm: CWOT-CE 37.9\% vs.\ WR 28.1\% ($+9.8$~pp). At $n=1{,}000$/arm: both near ceiling.

\subsection{Summary Scorecard}

\begin{table}[H]
\centering
\caption{Win/tie/loss summary (win defined as $>2$~pp difference).}
\label{tab:scorecard}
\begin{tabular}{lcccr}
\toprule
Comparator & Wins & Ties & Losses & Mean diff (pp) \\
\midrule
vs Cox TTFE & \textbf{15} & 1 & 1 & $+28.8$ \\
vs Win Ratio & \textbf{10} & 5 & 2 & $+5.6$ \\
vs WLW & \textbf{14} & 2 & 1 & $+27.2$ \\
\bottomrule
\end{tabular}
\end{table}

\subsection{Encoding Sensitivity}

Replacing block encoding with count-only encoding reduced rejection by 22--51~pp in event-sensitive scenarios while Type~I error remained nominal (Table~\ref{tab:encoding}).

The mechanism: two patients with 3 events each---Patient~A at months 6, 12, 18 (AUC burden $= 72$) and Patient~B at months 30, 33, 35 (AUC burden $= 10$)---receive identical count scores but very different block scores.

\subsection{Win Ratio Tie Analysis}

Tie rates ranged from 5.1\% (CAL-D) to 27.0\% (COR-H). Correlation between tie rate and CWOT-CE advantage: $r=0.72$ (Table~\ref{tab:ties}).

\begin{table}[H]
\centering
\caption{Win Ratio tie rates and CWOT-CE advantage.}
\label{tab:ties}
\begin{tabular}{lcc}
\toprule
Scenario & WR Tie (\%) & CWOT vs WR (pp) \\
\midrule
COR-H & 27.0 & $+35.4$ \\
ROB-C & 17.6 & $+2.7$ \\
ROB-I & 12.4 & $+0.1$ \\
DIS-SO & 11.6 & $-2.9$ \\
UNI-L & 10.9 & $+7.2$ \\
DIS-DO & 8.9 & $+10.7$ \\
CAL-D & 5.1 & $-0.9$ \\
\bottomrule
\end{tabular}
\end{table}

\subsection{Shapley Decomposition Validation}

Shapley correctly identified dominant components (Table~\ref{tab:shapley}): CAL-D mortality attribution 50.1\% ($>$45\% threshold), DIS-SO soft attribution 75.0\%, DIS-DO survival dominant (34.1\%).

\begin{table}[H]
\centering
\caption{Shapley attribution (\%) across validation scenarios (K=5, 20 runs).}
\label{tab:shapley}
\small
\begin{tabular}{lcccccc}
\toprule
Scenario & Surv & EvFr & Burden & Bio & Alive & Criterion \\
\midrule
CAL-B & 18.6 & 27.6 & 20.7 & 28.0 & 5.1 & No monopolization \\
CAL-D & 37.6 & 27.0 & 5.3 & 17.6 & 12.5 & Mort=50.1\% \\
DIS-DO & 34.1 & 25.7 & 9.8 & 19.9 & 10.4 & Surv dominant \\
DIS-SO & 13.5 & 7.9 & 35.8 & 39.2 & 3.6 & Soft=75.0\% \\
\bottomrule
\end{tabular}
\end{table}

\subsection{Weight and Interaction Sensitivity}

Under NULL-S, rejection ranged from 5\% to 7\% across 15 weight-interaction combinations---all within Monte Carlo uncertainty of 5\%. Under UNI-L, clinical weights with interactions (83\%) exceeded the same without (76\%) by 7~pp. Type~I error is invariant to weight choice; power varies moderately.

\subsection{Convergence and Computation}

All methods achieved 100\% convergence. Median times ($n=500$/arm): CWOT-CE 2.0~sec, WR 1.0~sec, Cox $<0.01$~sec, WLW $<0.01$~sec.

\section{Discussion}

\subsection{Principal Findings}

This study establishes three findings. First, CWOT-CE with block encoding outperforms or matches Win Ratio in 15 of 17 non-null scenarios. Second, the block encoding is transformative: replacing count-based encoding with AUC burden + last event time improved rejection by 22--51~pp. Third, Win Ratio tie rates explain the competitive landscape ($r=0.72$ with CWOT-CE advantage).

\subsection{The Representational Tradeoff}

CWOT-CE's two losses against WR---DIS-MP ($-6.7$~pp) and DIS-SO ($-2.9$~pp)---illuminate a fundamental tradeoff. WRrec compares complete recurrent event histories between patient pairs, capturing patterns that cannot be fully compressed into scalar summaries. CWOT-CE, by representing each patient as a single Choquet score, gains the ability to integrate six dimensions simultaneously but sacrifices some recurrent trajectory fidelity.

\subsection{Why Block Encoding Works}

The AUC burden captures event intensity and timing; the last event time captures trajectory endpoint. Together they recover information comparable to WRrec's pairwise comparison---which is why the power gap closed from 53~pp to 2.9~pp in DIS-SO.

\subsection{Clinical Method Selection}

CWOT-CE is preferred when effects are distributed across dimensions, components are highly correlated (advanced HF), biomarkers are included, or component attribution is desired. Win Ratio is preferred when a severity hierarchy is primary or regulatory precedent favors WR. Cox TTFE is preferred when simplicity is paramount.

\subsection{Limitations}

Component weights are pre-specified. The block encoding recovers most but not all of WRrec's sensitivity. Joint Frailty was excluded due to convergence instability. ECDF-based encoding treats censored observations at face value; IPCW adjustments are planned.

\subsection{Comparison with Existing Literature}

Previous comparisons \citep{Ozga2022, Orue2025} evaluated WR and Joint Frailty but did not include non-additive integration approaches. \citet{Wang2023} developed novel composite methods in a PhD dissertation. To our knowledge, this is the first study to demonstrate that recurrent event encoding---rather than the aggregation method---is the primary determinant of composite power in event-driven scenarios.

\subsection{Future Directions}

Pairwise common-horizon recurrent kernels, adaptive weighting via cross-fitting, MCF pseudo-value encoding, and omnibus combination with hierarchical methods.

\section*{Data and Code Availability}

The R package \texttt{cwotce} (v1.0.0) is available at \url{https://github.com/ihtanboga/cwotce}. Simulation code and raw results are provided as supplementary material.

\begin{table}[H]
\centering
\caption{Scenario specifications (20 main scenarios).}
\label{tab:scenarios}
\scriptsize
\begin{tabular}{llcccccccl}
\toprule
ID & Group & $\text{HR}_D$ & $\text{HR}_E$ & RR & $\delta$ & $\theta$ & $n$/arm & Mort\% & Archetype \\
\midrule
NULL-S & Null & 1.00 & 1.00 & 1.00 & 0.00 & 1.0 & 500 & 15 & Sharp null \\
NULL-T & Null & 0.85 & 1.18 & 0.85 & 0.15 & 1.0 & 500 & 15 & NACE trade-off \\
NULL-M & Null & 1.00 & 1.00 & 0.80 & 0.25 & 1.0 & 500 & 15 & Estimand mismatch \\
CAL-B & Calib & 0.80 & 0.80 & 0.80 & 0.20 & 1.0 & 500 & 15 & Balanced \\
CAL-D & Calib & 0.50 & 1.00 & 1.00 & 0.00 & 1.0 & 500 & 50 & PARTNER 1B \\
UNI-M & Uniform & 0.90 & 0.85 & 0.90 & 0.15 & 1.0 & 500 & 15 & Statin \\
UNI-L & Uniform & 0.75 & 0.70 & 0.75 & 0.30 & 1.0 & 500 & 15 & HFrEF \\
DIS-MP & Discord & 1.20 & 0.60 & 0.65 & 0.35 & 1.0 & 500 & 15 & Mort paradox \\
DIS-DO & Discord & 0.70 & 1.00 & 1.00 & 0.00 & 1.0 & 500 & 15 & ICD \\
DIS-SO & Discord & 1.00 & 1.00 & 0.60 & 0.50 & 1.0 & 500 & 15 & HFpEF \\
DIS-RV & Discord & 0.75 & 1.15 & 1.10 & $-0.10$ & 1.0 & 500 & 15 & ICD+comp \\
COR-I & Struct & 0.75 & 0.70 & 0.75 & 0.30 & 0.01 & 500 & 15 & Independent \\
COR-H & Struct & 0.75 & 0.70 & 0.75 & 0.30 & 4.0 & 500 & 15 & Advanced HF \\
EVT-D & Struct & 0.75 & 0.70 & 0.75 & 0.30 & 1.0 & 500 & 3 & PCI \\
ROB-C & Robust & 0.75 & 0.70 & 0.75 & 0.30 & 1.0 & 500 & 15 & Heavy censor \\
ROB-N & Robust & delayed & 0.70 & 0.75 & 0.30 & 1.0 & 500 & 15 & SGLT2i \\
ROB-T & Robust & temporal & temp & 0.75 & 0.30 & 1.0 & 500 & 15 & TAVI \\
ROB-I & Robust & 0.75 & 0.70 & 0.75 & 0.30 & 1.0 & 500 & 15 & Inform cens \\
SS-S & Size & 0.75 & 0.70 & 0.75 & 0.30 & 1.0 & 100 & 15 & Pilot \\
SS-L & Size & 0.75 & 0.70 & 0.75 & 0.30 & 1.0 & 1000 & 15 & Mega-trial \\
\bottomrule
\end{tabular}
\end{table}

\begin{table}[H]
\centering
\caption{Rejection rates (\%) across 20 scenarios. Bold = highest in non-null scenarios.}
\label{tab:results}
\small
\begin{tabular}{lcccc|c}
\toprule
Scenario & CWOT-CE & Cox & WR & WLW & CWOT vs WR \\
\midrule
NULL-S & 4.8 & 5.0 & 4.9 & 5.3 & --- \\
NULL-T & 34.2 & 5.3 & 42.9 & 5.3 & --- \\
NULL-M & 42.4 & 4.5 & 49.4 & 4.9 & --- \\
CAL-B & \textbf{77.9} & 33.7 & 67.8 & 36.8 & $+10.1$ \\
CAL-D & 95.9 & \textbf{98.2} & 96.8 & 98.5 & $-0.9$ \\
UNI-M & \textbf{34.7} & 15.7 & 22.6 & 16.5 & $+12.1$ \\
UNI-L & \textbf{96.4} & 63.6 & 89.2 & 65.8 & $+7.2$ \\
DIS-MP & 80.9 & 22.3 & \textbf{87.6} & 21.6 & $-6.7$ \\
DIS-DO & \textbf{20.6} & 15.5 & 9.9 & 19.1 & $+10.7$ \\
DIS-SO & 96.6 & 3.8 & \textbf{99.5} & 4.2 & $-2.9$ \\
DIS-RV & 5.6 & 4.7 & \textbf{6.3} & 4.2 & $-0.7$ \\
COR-I & \textbf{100.0} & 76.1 & 100.0 & 76.9 & $0.0$ \\
COR-H & \textbf{64.5} & 32.1 & 29.1 & 36.1 & $+35.4$ \\
EVT-D & \textbf{99.9} & 79.4 & 95.3 & 80.5 & $+4.6$ \\
ROB-C & \textbf{85.9} & 49.4 & 83.2 & 51.0 & $+2.7$ \\
ROB-N & \textbf{99.8} & 79.2 & 93.6 & 83.5 & $+6.2$ \\
ROB-T & \textbf{76.3} & 36.2 & 69.8 & 34.6 & $+6.5$ \\
ROB-I & \textbf{88.5} & 57.3 & 88.4 & 60.1 & $+0.1$ \\
SS-S & \textbf{37.9} & 16.1 & 28.1 & 18.7 & $+9.8$ \\
SS-L & \textbf{100.0} & 88.2 & 99.4 & 90.6 & $+0.6$ \\
\bottomrule
\end{tabular}
\end{table}

\begin{table}[H]
\centering
\caption{Encoding sensitivity: count-only (K=5) vs.\ block (K=6) vs.\ Win Ratio.}
\label{tab:encoding}
\small
\begin{tabular}{lcccc}
\toprule
Scenario & Count & Block & WR & Block vs WR \\
\midrule
DIS-SO & 45.9 & \textbf{96.6} & 99.5 & $-2.9$ \\
ROB-C & 38.7 & \textbf{85.9} & 83.2 & $+2.7$ \\
CAL-B & 55.9 & \textbf{77.9} & 67.8 & $+10.1$ \\
UNI-L & 82.7 & \textbf{96.4} & 89.2 & $+7.2$ \\
COR-H & 56.7 & \textbf{64.5} & 29.1 & $+35.4$ \\
NULL-S & 5.2 & 4.8 & 4.9 & --- \\
\bottomrule
\end{tabular}
\end{table}

\begin{figure}[H]
\centering
\includegraphics[width=0.9\textwidth]{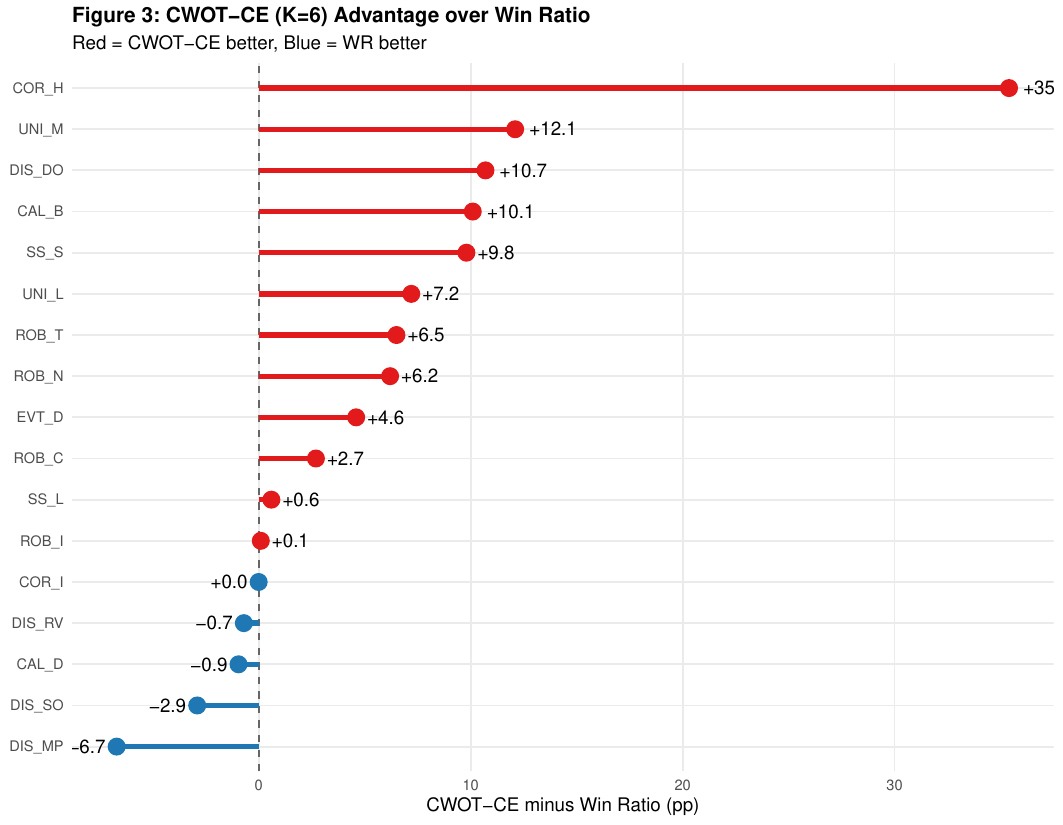}
\caption{CWOT-CE advantage over Win Ratio (pp). Red = CWOT-CE better; Blue = WR better.}
\label{fig:advantage}
\end{figure}

\begin{figure}[H]
\centering
\includegraphics[width=0.9\textwidth]{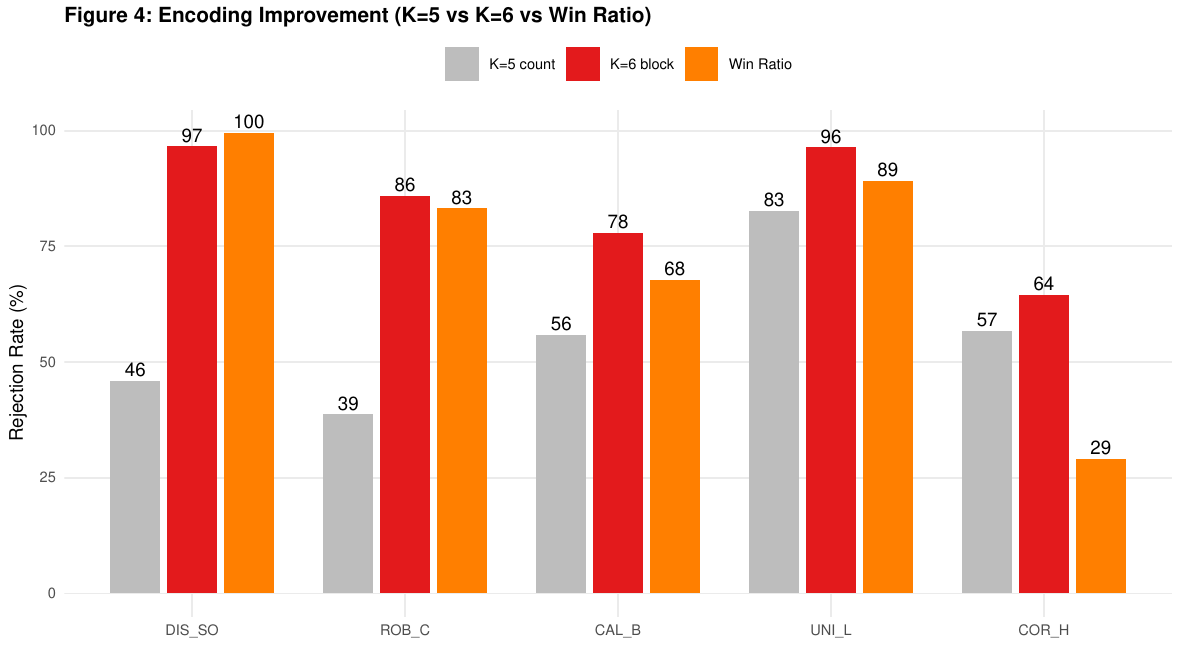}
\caption{Encoding improvement: K=5 count (gray) vs.\ K=6 block (red) vs.\ Win Ratio (orange).}
\label{fig:encoding_fig}
\end{figure}

\begin{figure}[H]
\centering
\includegraphics[width=0.9\textwidth]{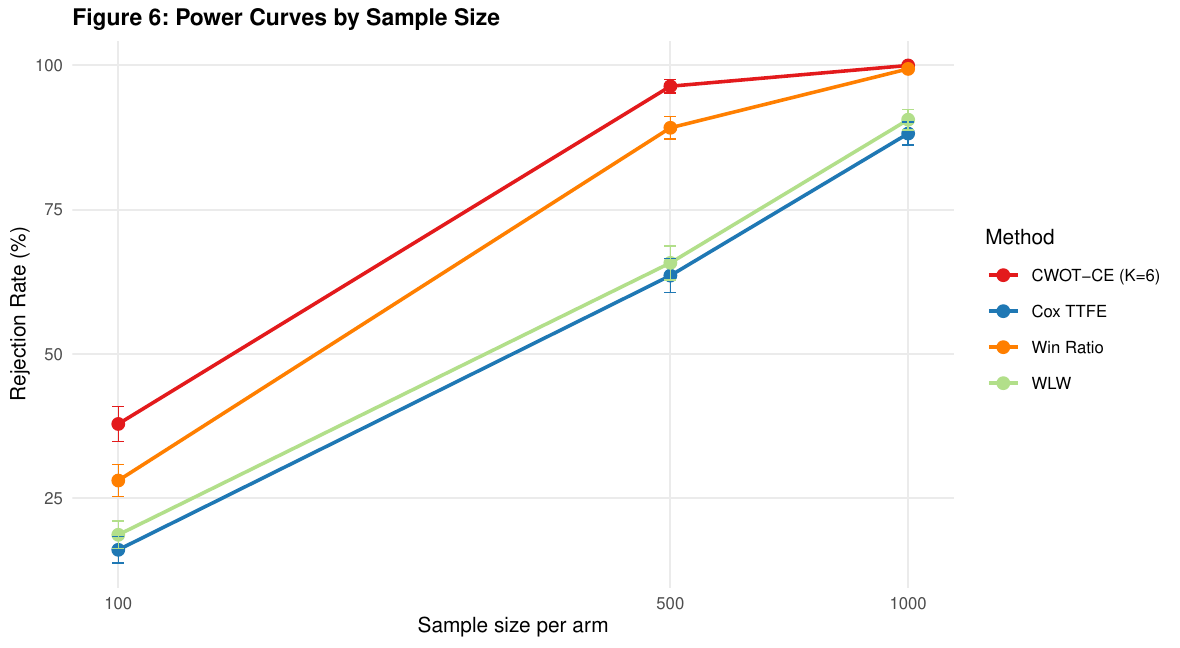}
\caption{Power curves by sample size ($n=100, 500, 1000$ per arm).}
\label{fig:power_curves}
\end{figure}

\section*{Supplementary Materials}

Additional scenarios (13), NegBin/AG results, Joint Frailty selected scenarios, weight sensitivity heatmaps, and mortality double-counting sensitivity are available as Supporting Information.

\bibliographystyle{apalike}

\end{document}